\documentclass[aps,prc,superscriptaddress,showpacs,floatfix,nofootinbib,notitlepage]{revtex4-1}
\usepackage{amsmath,graphicx,float,latexsym,hyperref,subfigure,bbold}

\begin{document}
\title{Forward di-hadron back-to-back correlations in $\boldsymbol{pA}$ collisions from rcBK evolution}

\author{Javier L. Albacete}
\affiliation{CAFPE and Departamento de F\'isica Te\'orica y del Cosmos, Universidad de Granada, E-18071 Campus de Fuentenueva, Granada, Spain}
\author{Giuliano Giacalone}
\affiliation{Institut de physique th\'eorique, Universit\'e Paris Saclay, CNRS,
CEA, F-91191 Gif-sur-Yvette, France}
\affiliation{CPHT, \'Ecole Polytechnique, CNRS, Universit\'e Paris Saclay, Route de Saclay, 91128 Palaiseau, France}
\author{Cyrille Marquet}
\affiliation{CPHT, \'Ecole Polytechnique, CNRS, Universit\'e Paris Saclay, Route de Saclay, 91128 Palaiseau, France}
\author{Marek Matas}
\affiliation{Czech Technical University in Prague, FNSPE, B\v{r}ehov\'a 7, 11519 Prague, Czech Republic}

\begin{abstract}
We study the disappearance of the away-side peak of the di-hadron correlation function in p+A vs p+p collisions at forward rapidities, when the scaterring process presents a manifest dilute-dense asymmetry. We improve the state-of-the-art description of this phenomenon in the framework of the Color Glass Condensate (CGC), for hadrons produced nearly back-to-back. In that case, the gluon content of the saturated nuclear target can be described with transverse-momentum-dependent gluon distributions, whose small-$x$ evolution we calculate numerically by solving the Balitsky-Kovchegov equation with running coupling corrections.
We first show that our formalism provides a good description of the disappearance of the away-side azimuthal correlations in d+Au collisions observed at BNL Relativistic Heavy Ion Collider (RHIC) energies. Then, we predict the away-side peak of upcoming p+Au data at $~\sqrt[]{s}=200$ GeV to be suppressed by about a factor 2 with respect to p+p collisions, and we propose to study the rapidity dependence of that suppression as a complementary strong evidence of gluon saturation in experimental data.
\end{abstract}

\maketitle

\section{Introduction}
\label{sec:1}
Azimuthal correlations of particles in the final states of hadronic collisions serve as a powerful tool for experimental tests of the Color Glass Condensate (CGC) \cite{Gelis:2010nm,Gelis:2012ri,Blaizot:2016qgz}, the effective theory of protons and nuclei in the nonlinear regime of quantum chromodynamics. 
A special role in the phenomenology of the CGC is played by correlations of particles in p+A collisions probed in the region of fragmentation of the protons~\cite{Marquet:2007vb,Tuchin:2009nf,Albacete:2010pg,Stasto:2011ru,Lappi:2012nh,Ayala:2016lhd,vanHameren:2016ftb,Kotko:2017oxg}, where the rapidities of the correlated particles are large and positive (forward rapidity region).
Such configurations are ideal for testing the CGC theory, because they induce a dilute-dense asymmetry in the problem: The projectile proton is probed at large values of Bjorken $x$, and is thus a dilute object, amenable to a description in terms of well-known parton distribution functions (PDFs); The nuclear target is instead seen as dense state of low $x$ gluons, a regime in which the \textit{saturation} of the gluon densities is manifest, so that the CGC description applies.
This dilute-dense asymmetry hence minimizes our uncertainty in the knowledge of the projectile, and provides the cleanest possible environment for the study of phenomenological signatures of gluon saturation in the target. 

In this paper, we deal with a salient prediction of the CGC theory: The disappearance of the away-side peak ($\Delta\phi=\pi$) of the two-particle correlation function of dilute-dense collisions (i.e., forward p+A collisions).
Following \cite{Albacete:2014fwa}, let us provide an intuitive picture of this phenomenon.
A valence parton interacting with a CGC (i.e., a large classical Yang-Mills background field) undergoes multiple scattering with low-$x$ gluons, either before or after splitting into a pair of back-to-back partons, which eventually produce the jets or hadrons observed in the final state.
The pair of partons is put on shell via the interaction with the target, and this occurs through a transverse momentum exchange of order of the saturation scale of the target, $Q_s$, which is typically much larger than the transverse momentum of the parent valence parton. The back-to-back correlation of the final-state particles, which would not be affected by an interaction with a gluon of zero transverse momentum, is therefore altered, and this induces a depletion of the correlation function around $\Delta\phi=\pi$.
Consequently, the away-side peak observed in p+A collisions is expected to be suppressed with respect to that of p+p collisions, because the target nuclei are denser and more saturated.

Experimentally, the validity of this picture is strongly supported by Relativistic Heavy-Ion Collider (RHIC) data, as both the STAR and the PHENIX collaborations reported a visible suppression of the away-side peak when comparing p+p collisions to central d+Au collisions~\cite{Braidot:2010zh,Adare:2011sc}.
These data, though, suffer from large uncertainties.
More accurate tests of the CGC prediction may nevertheless become possible with the advent of data from the recent 200 GeV p+Au run performed at RHIC.
As we shall see, one of the goal of this paper is to provide predictions for the away-side peak in these collisions.

On the theory side, first calculations of forward two-particle production in p+A collisions within the CGC framework date back to more than 10 yrs ago \cite{JalilianMarian:2004da,Marquet:2007vb}.
The cross section for the production of two particles is intrinsically difficult to evaluate, because it involves multi-point correlators of Wilson lines.
Over the years, different levels of approximation have been employed to perform calculations and obtain predictions, as reviewed in Ref.~\cite{Albacete:2014fwa}. The simplest option is to disregard non-linear effects and recover the so-called $k_t-$factorization (or high-energy factorization) framework \cite{Tuchin:2009nf,Kutak:2012rf}; the cross-section is then obtained from a single two-point correlator, but that approximation is not applicable in the away-side peak region. In Ref.~\cite{Albacete:2010pg}, the multi-point correlators are evaluated using the so-called Gaussian approximation of the non-linear QCD evolution, however only the {\it elastic} contributions are kept, and it turns out that the neglected contributions are also sizable in the away-side peak region. In Ref.~\cite{Lappi:2012nh}, the complete Gaussian expressions are used, however due to the complexity of the problem, only quark-initiated channels could be included. 

A crucial step was the realization that the cross section simplifies dramatically if one considers the production of partons which are nearly back-to-back~\cite{Dominguez:2010xd,Dominguez:2011wm}.
In this limit, the dense part of the scattering (the nucleus) is characterized by transverse-momentum-dependent (TMD) gluon distributions whose small $x$ evolution is easily affordable to numerical implementations, because the multi-point correlators of Wilson line~\cite{Marquet:2016cgx} involve only two distinct transverse positions. 
This framework has been employed in a number of applications~\cite{Stasto:2011ru,vanHameren:2016ftb,Marquet:2016cgx,Marquet:2017xwy,Kotko:2017oxg,Boer:2017xpy,Marquet:2017xwy,Altinoluk:2018uax}, and was recently reviewed in \cite{Petreska:2018cbf}\footnote{Note that an improved version of the TMD formalism (dubbed ITMD) was introduced in Ref.~\cite{Kotko:2015ura}. This framework allows one to relax the condition $\Delta\phi\sim\pi$, and was applied in calculations of forward di-jet production in Refs.~ \cite{vanHameren:2016ftb,Kotko:2017oxg}. However, in this paper we do not employ the improved framework because it is strictly equivalent to the original TMD formulation as long as back-to-back particles are considered.}. In the case of forward di-hadron production, it has only been used together with Golec-Biernat Wusthoff (GBW) type parameterizations for the gluon TMDs \cite{Stasto:2011ru}, which suffer from unphysical exponential tails at large gluon transverse momentum. The goal of our work is to improve on this by obtaining the TMD gluon distributions from numerical solutions of the QCD non-linear evolution.

In \cite{Marquet:2016cgx,Marquet:2017xwy}, the gluon TMDs and their small-$x$ evolution were obtained from the full QCD evolution at leading logarithmic accuracy, i.e. from the Jalilian-Marian--Iancu--McLerran--Weigert--Leonidov--Kovner (JIMWLK) equation. 
Since the implementation of running-coupling corrections in this context has not been performed yet, we prefer to work within the Gaussian approximation of JIMWLK evolution and obtain the gluon TMDs from the Balitsky-Kovchegov (BK) equation with running coupling corrections (rcBK), because we expect running-coupling corrections to be much more important than corrections to the Gaussian approximation. 
In addition, the rcBK solutions are well constrained from deep inelastic scattering data \cite{Albacete:2010sy}, so that the final expression of the cross section of two-parton production turns out to be essentially free from tunable parameters.
We shall derive this cross section, and convolute it with fragmentation functions to present state-of-the-art results on azimuthal correlations of di-hadrons in forward p+A and p+p collisions at RHIC energies. We both test our theory against existing data and make predictions for future back-to-back correlations of hadrons.

The paper is organized as follows.
In Sec.~\ref{sec:2}, we briefly review the theoretical formalism of nearly-back-to-back forward di-hadron production in p+A collisions in the CGC framework, and we present the fully differential cross section for the production of di-hadrons, specifically, two neutral pions.
In the cross section we shall introduce the TMD gluon distributions which characterize the dense component of the scattering process. 
In Sec.~\ref{sec:3}, we explain in detail how such quantities are obtained from rcBK evolution, and we show their behavior as function of the kinematic variables.
Calculations of the away-side peak are eventually given in Sec.~\ref{sec:4}.
The per-trigger-yield cross section is calculated as function of the relative azimuthal angle of the two hadrons.
We first calculate it in d+Au and p+p collisions, and we compare our results to existing RHIC data.
Then, we compute several predictions for the away-side peak in upcoming p+Au collisions at $~\sqrt[]{s}=200$~GeV.
In Sec.~\ref{sec:5}, we make predictions for the evolution with rapidity of the suppression of the away-side peak, and we compare our results with calculations performed using an alternative implementation of the rcBK evolution for nuclei.
Section~\ref{sec:6} is left for conclusive remarks.

\section{Color Glass Condensate in the back-to-back region: TMD factorization}
\label{sec:2}
\begin{figure}[t]
\centering
\includegraphics[width=.4\linewidth]{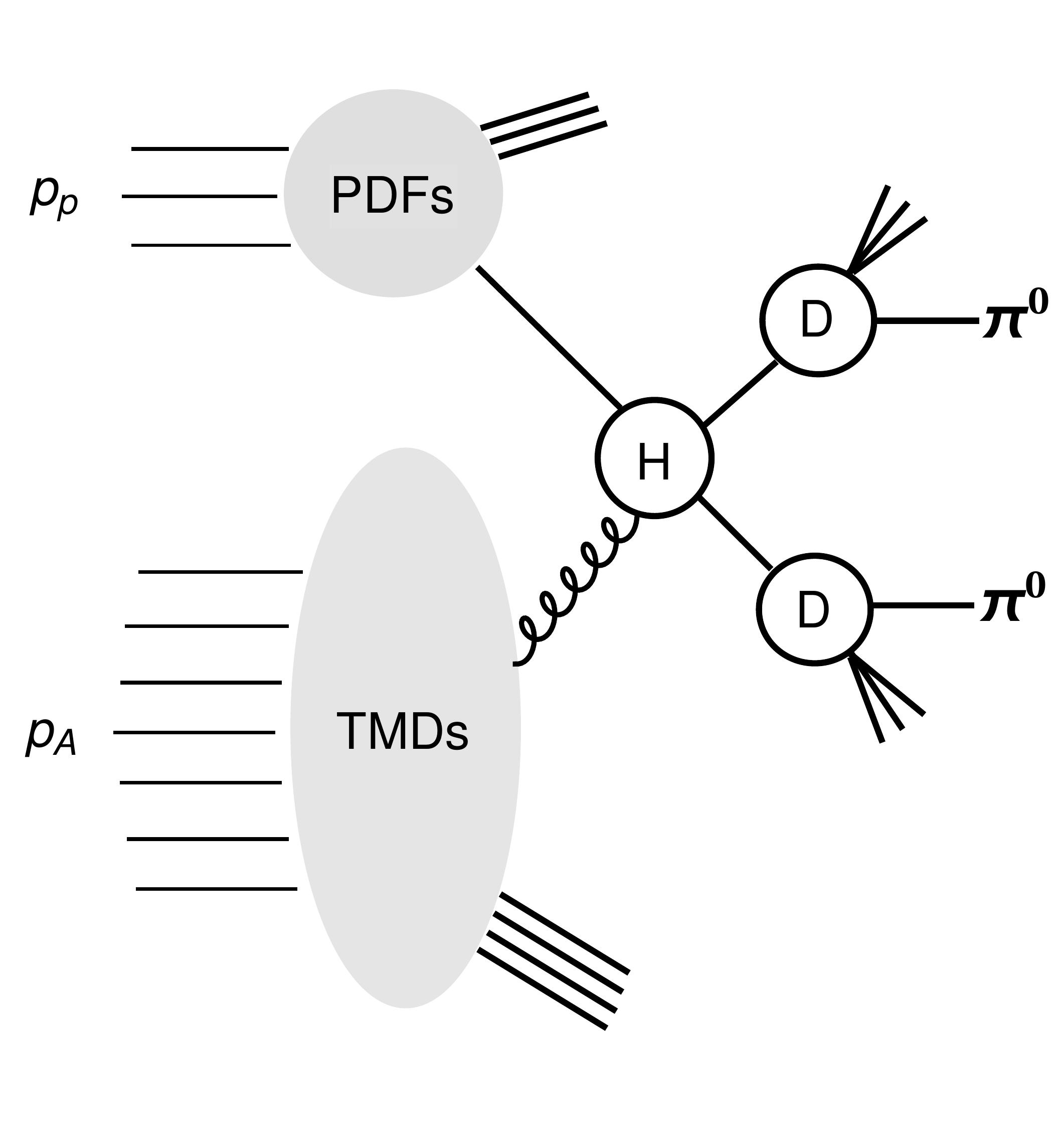}
\caption{The $pA \rightarrow \pi^0\pi^0X$ process. See the text for details about the displayed quantities.}
\label{fig:1}
\end{figure}

We study the production of pairs of hadrons in forward p+A collisions. 
We display such process in Fig.~\ref{fig:1}.
Working in light-cone coordinates, ({\small +},~${\small \perp}$,~$-$), the Feynman-$x$ variables associated with the projectile parton moving along the {\small +} direction and with the target gluon coming from the $-$ direction are, respectively, given by 
\begin{align}
\label{eq:kine}
\nonumber x_1 &= \frac{k_1^++k_2^+}{p_p^+}=\frac{1}{\sqrt[]{s}}\Big(\frac{p_{1t}}{z_1}~e^{y_1}+\frac{p_{2t}}{z_2}~e^{y_2}\Big), \\
x_2 &= \frac{k_1^-+k_2^-}{p_A^-}=\frac{1}{\sqrt[]{s}}\Big(\frac{p_{1t}}{z_1}~e^{-y_1}+\frac{p_{2t}}{z_2}~e^{-y_2}\Big),
\end{align}
where the $k_i$'s refer to the outgoing partons while the $p_{i}'s$ refer to the final-state hadrons. We have introduced their transverse momenta, $p_{1t}\!=\!z_{1}k_{1t}$ and $p_{2t}\!=\!z_{2}k_{2t}$, and rapidities, $y_1$ and $y_2$, while $~\sqrt[]{s}$ denotes the invariant mass of the scattering process.
Equation~(\ref{eq:kine}) shows that when both $y_1$ and $y_2$ are large and positive, we probe a large-$x$ parton in the projectile, and a small-$x$ gluon inside the target nucleus. 
In experiments at RHIC, one can reach $y\sim4$, leading to $x_1\sim0.5$, and $x_2\sim10^{-3}$: This realizes the anticipated dilute-dense asymmetry of forward particle production, which is essential for the applicability of our formalism.

Now, following \cite{Dominguez:2011wm,Marquet:2016cgx}, we dub 
\begin{equation}
z=\frac{k_1^+}{k_1^+ + k_2^+}=\frac{p_1^+/z_1}{p_1^+/z_1 + p_2^+/z_2},
\end{equation}
and we introduce the following variables
\begin{equation}
k_t = k_{1t} + k_{2t}, \hspace{40pt} P_t=(1-z)k_{1t}-zk_{2t}.
\end{equation}
If we stick to a limit in which the produced particles are back-to-back, i.e., their relative azimuthal angle, $\Delta\phi$, is close to $\pi$, then the total transverse momentum of the di-hadron pair is much smaller than the transverse momentum of the single hadrons, i.e., $|k_t| \ll |P_t|$~\cite{Dominguez:2011wm}.

Following the exhaustive derivations of \cite{Marquet:2016cgx}, the advantage of this limit is that it allows to write the cross section of the scattering process as an expansion in powers of $1/P_t$.
Keeping only the leading order terms in this expansion, the dense component of the scattering is given by a combination of transverse momentum dependent gluon distributions (TMDs in short), which are CGC correlators of traces of Wilson lines.
Summing over all production channels ($qg \rightarrow qg$, $gg \rightarrow q\bar q$, $gg \rightarrow gg$), the cross section for the production of two partons can be written in the following compact notation \cite{Marquet:2016cgx}
\begin{equation}
\label{eq:tmdf}
\frac{ d \sigma^{pA\rightarrow hh X}}{ d^2 k_{1t}  d^2 k_{2t}  d y_1  d y_2} = \frac{\alpha_s^2}{(x_1 x_2 s)^2} \sum_{a,c,d} x_1 f_{a/p} (x_1,\mu^2) \sum_{i} \frac{1}{1+\delta_{cd}} H_{ag\rightarrow cd}^{(i)}(z,P_t)\mathcal{F}_{ag}^{(i)}(x_2,k_t),
\end{equation}
where we note the manifest factorization of the cross section into a dilute component, characterized by collinear parton distribution functions $f_{a/p}(x_1,\mu^2)$, evaluated at a factorization scale $\mu^2$, and a channel-dependent dense component, characterized by hard factors~\cite{Dominguez:2011wm}, $H(z,P_t)$, and the TMD gluon distributions, $\mathcal{F}^{(i)}(x_2,k_t)$, specified below.

To turn Eq.~(\ref{eq:tmdf}) into a tool enabling us to compute predictions for di-hadron production, we convolute it with fragmentation functions.
Considering $u$ quarks, $d$ quarks and gluons in the projectile proton, and considering only their fragmentation into pions, and neglecting all terms which are suppressed by $1/N_c^2$, the full expression of the cross section reads
\begin{align}
\label{eq:full}
&\nonumber\frac{d\sigma^{pA\rightarrow \pi^0\pi^0 X}}{dy_1~ dy_2~ d^2p_{1t}~ d^2p_{2t}} = \frac{\alpha_s^2}{2 C_F} \int_{p_{t1}\frac{e^{y_1}}{\sqrt[]{s}}/(1-p_{t2}\frac{e^{y_2}}{\sqrt[]{s}})}^1 \frac{dz_1}{z_1^2}   \int_{p_{t2}\frac{e^{y_2}}{\sqrt[]{s}}/(1-\frac{p_{t1}}{z_1}\frac{e^{y_1}}{\sqrt[]{s}})}^1 \frac{dz_2}{z_2^2} ~~ \frac{z(1-z)}{P_t^4}\\
\nonumber & \biggl\{ D_{\pi^0/g}(z_1,\mu^2) \bigl[ x_1 u(x_1,\mu^2)~D_{\pi^0/u}(z_2,\mu^2)+x_1d(x_1,\mu^2)~D_{\pi^0/d}(z_2,\mu^2) \bigr] P_{gq}(z)~\times \\
\nonumber &~~~~~~~~~~~\times~\biggl[ (1-z)^2 \mathcal{F}_{qg}^{(1)}(x_2,k_t) + \mathcal{F}_{qg}^{(2)}(x_2,k_t) \biggr]+\\
\nonumber & +~ D_{\pi^0/g}(z_2,\mu^2) \bigl[ x_1 u(x_1,\mu^2)~D_{\pi^0/u}(z_1,\mu^2)+x_1d(x_1,\mu^2)~D_{\pi^0/d}(z_1,\mu^2) \bigr] P_{gq}(1-z)~\times \\
\nonumber &~~~~~~~~~~~\times~\biggl[ z^2 \mathcal{F}_{qg}^{(1)}(x_2,k_t) + \mathcal{F}_{qg}^{(2)}(x_2,k_t) \biggr]+\\
\nonumber& + ~ 2 \bigl[ D_{\pi^0/u}(z_1,\mu^2)~D_{\pi^0/u}(z_2,\mu^2) + D_{\pi^0/d}(z_1,\mu^2)~D_{\pi^0/d}(z_2,\mu^2) \bigr]~x_1g(x_1,\mu^2)~P_{qg}(z)~\times \\
\nonumber &~~~~~~~~~~~\times~\biggl[ \mathcal{F}_{gg}^{(1)}(x_2,k_t) - 2z(1-z) \bigl(\mathcal{F}_{gg}^{(1)}(x_2,k_t) - \mathcal{F}_{gg}^{(2)}(x_2,k_t)\bigr) \biggr] + \\
\nonumber & + ~ D_{\pi^0/g}(z_1,\mu^2) D_{\pi^0/g}(z_2,\mu^2)~x_1g(x_1,\mu^2)~P_{gg}(z)~\times \\
&~~~~~~~~~~~\times~\biggl[ \mathcal{F}_{gg}^{(1)}(x_2,k_t) - 2z(1-z) \bigl(\mathcal{F}_{gg}^{(1)}(x_2,k_t) - \mathcal{F}_{gg}^{(2)}(x_2,k_t)\bigr) + \mathcal{F}_{gg}^{(6)}(x_2,k_t) \biggr]~ \biggr\},
\end{align}
where we have denoted by $D_{\pi^0/a}(z_i,\mu^2)$ the fragmentation of a parton $a$ into a neutral pion at the factorization scale $\mu^2$, and the notation used for the distributions $\mathcal{F}^{(i)}_{ag}(x_2,k_t)$ is the same as in~\cite{Marquet:2016cgx}.
In Sec.~\ref{sec:4} we shall make use of Eq.~(\ref{eq:full}) to compute azimuthal correlations of neutral pions at RHIC.
Let us first describe, in the following section, the rcBK formalism developed for the small $x_2$ evolution of the TMD distributions which appear in the cross section.

\section{Evolution of the TMD gluon distributions towards small $\boldsymbol{x}$}
\label{sec:3}

In order to complete our formulation of the cross section, Eq.~(\ref{eq:full}), we discuss now the $x_2$ evolution of the TMD gluon distributions, $\mathcal{F}_{ag}^{(i)}(x_2,k_t)$. 
The starting point is the evolution of the impact parameter ($b$) independent fundamental--dipole scattering amplitude, which we denote in a standard notation $N_F(x,r)$. 
As it is customarily done in the literature, we assume that the $b$ dependence of $N_F$ factorizes, and that it does not mix with the evolution.
The evolution equation of the dipole amplitude, known as the Balitsky-Kovchegov equation \cite{Balitsky:1995ub,Kovchegov:1999yj}, supplemented with running coupling corrections (rcBK equation), reads ($r_i=|\bf r_i|$)
\begin{eqnarray}
  \frac{\partial N_{F}(r,x)}{\partial\ln(x_0/x)}=\int d^2r_1\
  K^{{\rm run}}({\bf r},{\bf r_1},{\bf r_2})
  \left[N_{F}(r_1,x)+N_{F}(r_2,x)
-N_{F}(r,x)\right.\nonumber\\
\left. -N_{F}(r_1,x)\,N_{F}(r_2,x)\right]\ ,
\label{bk1}
\end{eqnarray}
with ${\bf r_2}\equiv{\bf r}-{\bf r_1}$, and where $x_0$ is some initial value
for the evolution (usually chosen to be $x_0=0.01$). ${K}^{\rm run}$ is the evolution
kernel including running coupling corrections. Different prescriptions have been
proposed in the literature for $K^{\rm run}$. As shown in \cite{Albacete:2007yr}, Balitsky's prescription minimizes the role of
higher {\it conformal} corrections:
\begin{equation}
  K^{{\rm run}}({\bf r},{\bf r_1},{\bf r_2})=\frac{N_c\,\alpha_s(r^2)}{2\pi^2}
  \left[\frac{1}{r_1^2}\left(\frac{\alpha_s(r_1^2)}{\alpha_s(r_2^2)}-1\right)+
    \frac{r^2}{r_1^2\,r_2^2}+\frac{1}{r_2^2}\left(\frac{\alpha_s(r_2^2)}{\alpha_s(r_1^2)}-1\right) \right]\,.
\label{kbal}
\end{equation}

The rcBK evolution is independent of whether the target is a proton or a nucleus. That is accounted for in the initial condition. 
We use the so-called McLerran-Venugopalan (MV) model:
\begin{equation}
N_F(r,x\!=\!x_0)=
1-\exp\left[-\frac{r^2\,Q_{s0}^2}{4}\,
  \ln\left(\frac{1}{\Lambda\,r}+e\right)\right]\ ,
\label{ic}
\end{equation}
with $\Lambda=0.241$~GeV, and where $Q_{s0}$ denotes the saturation scale at the initial value $x_{0}$. We use $x_0=0.01$ and $Q_{s0}^2=0.2$ GeV$^2$ for a proton target, which are known to provide a good description of single-inclusive forward hadron RHIC data \cite{Albacete:2010bs}. For a target nucleus, things are more uncertain, as we are interested only in central collisions, i.e., collisions at small impact parameter. Motivated by previous studies~\cite{Albacete:2010pg}, we keep $x_0=0.01$, and we choose $Q_{s0}^2=0.6$~GeV$^2$, i.e., a factor 3 larger than the $Q_{s0}^2$ with a target proton.  

Now, the simplest gluon TMD distribution, ${\cal F}_{qg}^{(1)}(x_2,k_t)$, is related to the Fourier transform of the fundamental dipole amplitude, $N_F(x_2,{\bf{r}})$, and is given by:
\begin{eqnarray}
{\cal F}_{qg}^{(1)}(x_2,k_t)&=& \frac{N_c}{\alpha_s \pi (2\pi)^3}\int d^2b\int d^2 {\bf{r}}\
e^{-i k_t \cdot{\bf r}}\nabla^2_{{\bf{r}}}\ N_F(x_2, {\bf{r}})\nonumber\\
&=&\frac{N_c\ k_t^2\ S_\perp}{2\pi^2 \alpha_s} F(x_2,k_t) \ ,
\label{eq:dipgluontmd}
\end{eqnarray}
where
\begin{equation} 
F(x_2,k_t) = \int \frac{d^2{\bf{r}}}{(2\pi)^2} e^{-i{k_t}\cdot{\bf r}}[1-N_F(x_2, {\bf{r}})] ,
\end{equation}
and with $S_\perp$ denoting the transverse area of the target.

In full generality, none of the other gluon TMDs can be obtained in such a straightforward manner, directly from $N_F$, or its Fourier transform $F$. To move forward, we resort to a mean-field type approximation: we shall utilize the so-called Gaussian approximation of the CGC
\cite{Fujii:2006ab,Marquet:2007vb,Kovchegov:2008mk,Marquet:2010cf,Dumitru:2011vk,Iancu:2011nj,Alvioli:2012ba}. The essence of this approximation is to consider all the color charge correlations in the target to stay Gaussian throughout the evolution. This approximation, along with the large $N_c$-limit, ensures the factorization of CGC expectation values into single-trace expectation values, and allows to calculate ${\cal F}_{gg}^{(1)}$ and ${\cal F}_{gg}^{(2)}$ from $F$ \cite{Dominguez:2011wm}:
\begin{eqnarray}
  {\cal F}_{gg}^{(1)}(x_2,k_t)
  & = &
  \int d^2q_t\,{\cal F}_{qg}^{(1)}(x_2,q_t) F(x_2, k_t-q_t)\,,
  \label{GeneralFgg1}
  \\
  {\cal F}_{gg}^{(2)}(x_2,k_t)
  & = &
  -\int d^2q_t\frac{(k_t-q_t)\cdot q_t}{q_t^2}
  {\cal F}_{qg}^{(1)}(x_2,q_t) F(x_2,k_t-q_t)\,,
  \label{GeneralFgg2}
\end{eqnarray}
We note that the difference
\begin{eqnarray}
{\cal F}_{gg}^{(1)}(x_2,k_t)-{\cal F}_{gg}^{(2)}(x_2,k_t) &=&
\int d^2q_t\frac{k_t\cdot q_t}{q_t^2} {\cal F}_{qg}^{(1)}(x_2,q_t) F(x_2, k_t-q_t)\\&=&
\frac{k_t^2}{2} \int \frac{d^2q_t}{q_t^2} {\cal F}_{qg}^{(1)}(x_2,q_t) F(x_2, k_t-q_t)\label{Fadj}\ ,
\end{eqnarray}
which enters the cross-section \eqref{eq:full}, is related to the adjoint-dipole scattering amplitude $N_A$ in the same way that ${\cal F}_{qg}^{(1)}$ was related to the fundamental dipole scattering amplitude. Indeed, if we introduce
\begin{eqnarray}
{\cal F}_{\rm adj}(x_2,k_t)&=& \frac{C_F}{\alpha_s \pi (2\pi)^3}\int d^2b\int d^2 {\bf{r}}\
e^{-i k_t \cdot{\bf r}}\nabla^2_{{\bf{r}}}\ N_A(x_2, {\bf{r}})\nonumber\\
&=&\frac{C_F\ k_t^2\ S_\perp}{2\pi^2 \alpha_s} \tilde{F}(x_2,k_t) \ ,
\label{eq:adjdipgluontmd}
\end{eqnarray}
where
\begin{equation} 
\tilde{F}(x_2,k_t) = \int \frac{d^2{\bf{r}}}{(2\pi)^2} e^{-i{k_t}\cdot{\bf r}}[1-N_A(x_2, {\bf{r}})]\ ,
\end{equation}
then one has ${\cal F}_{gg}^{(1)}-{\cal F}_{gg}^{(2)}={\cal F}_{\rm adj}$. This identify is true in full generality, beyond the Gaussian and large-$N_c$ approximations (for which $1-N_A=[1-N_F]^2$) used here, as was first noticed in \cite{Marquet:2017xwy}.

Finally, the two remaining gluon TMDs need to be computed from the Weizs{\"a}cker-Williams (WW) gluon distribution~\cite{Dominguez:2011wm}, which we denote ${\cal F}_{WW}$, and which should be obtained from the quadrupole operator $\left< \text{Tr} \left [ A({\bf x})A({\bf y})\right]\right>_{x_2}$ where $A({\bf x})=U^\dagger({\bf x})\partial_{\bf x}U({\bf x})$ with $U$ denoting a Wilson line. Again, this quantity is in general not related to the solution of the BK equation, $F(x_2,k_t)$, but using the Gaussian approximation one can write (in the large $N_c$ limit)\footnote{Strickly speaking, the $1/r^2$ factor - which we shall keep in our numerical computation - should be replaced by a more complicated function of $N_F$, equal to $1/r^2$ only in the MV model at $x_2=x_0$.}:
\begin{equation}
\mathcal{F}_{WW}(x_2,k_t)= \frac{C_F}{2 \alpha_s \pi^4}\int d^2b\int \frac{d^2{\bf r}}{r^2}\
e^{-i{k_t}\cdot{\bf r}}\ \left\{1 - [1-N_F(x_2,{\bf r})]^2\right\} \ .
\label{eq:WW_AdjointDipole}
\end{equation}
which allows to calculate the remaining two gluon TMDs needed in the cross section as follows \cite{Dominguez:2011wm}: 
\begin{eqnarray}
  {\cal F}_{qg}^{(2)}(x_2,k_t)
  & = &
  \int d^2q_t\,{\cal F}_{WW}(x_2,q_t) F(x_2, k_t-q_t)\,,
  \\
  {\cal F}_{gg}^{(6)}(x_2,k_t)
  & = &
  \int d^2q_td^2q_t'\,{\cal F}_{WW}(x_2,q_t) F(x_2,q_t') F(x_2,k_t-q_t-q_t')\ .
  \label{GeneralFgg6}
\end{eqnarray}
We have now expressed all the needed gluon TMDs in terms of $F(x_2,k_t)$, the solution of the BK equation.
\begin{figure}[t!]
\centering
\includegraphics[width=.95\linewidth]{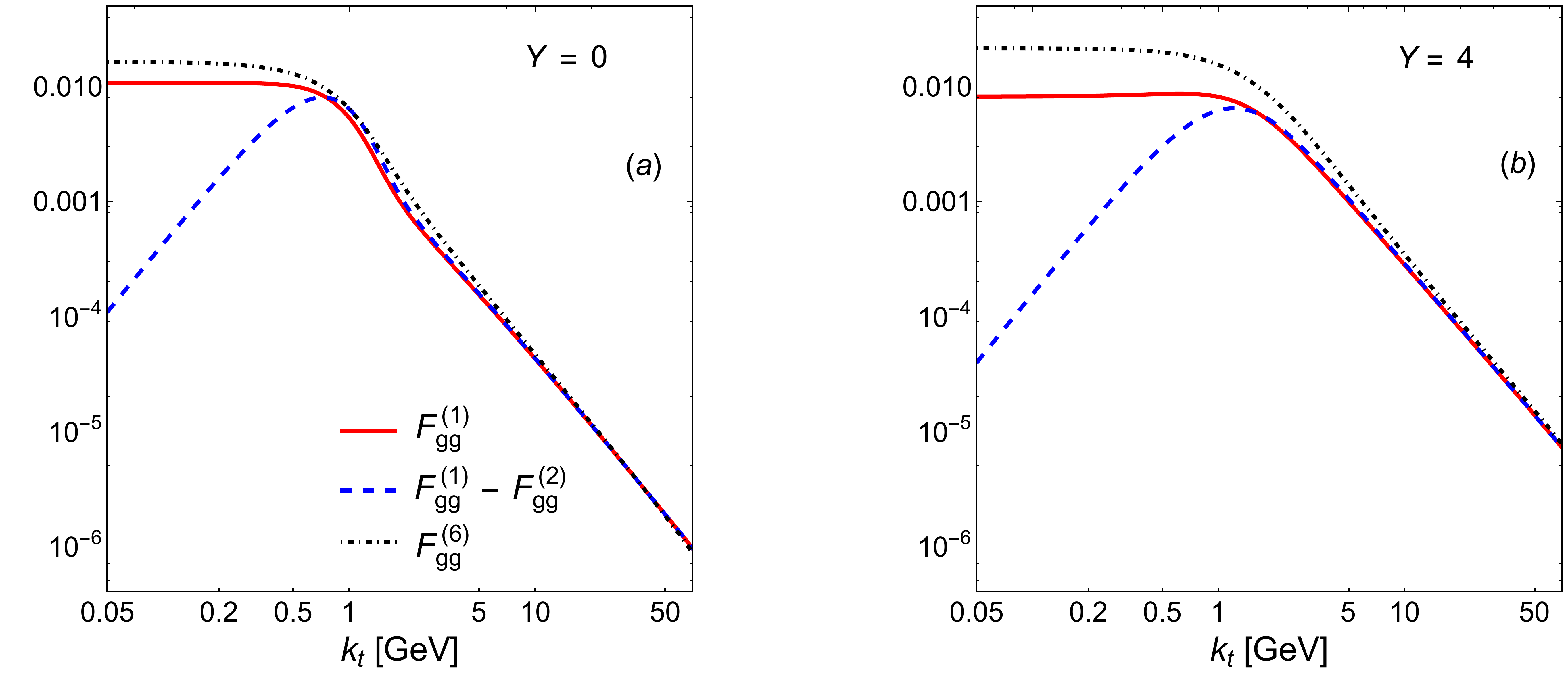}
\caption{This figure presents the $x_2$ evolution of three TMDs appearing in the cross section of Eq.~(\ref{eq:full}), for a target proton. In panel (a), the initial conditions at $x_2=0.01$ are presented. We show $\mathcal{F}_{gg}^{(1)}$ (solid line), $\mathcal{F}_{gg}^{(1)}-\mathcal{F}_{gg}^{(2)}$ (dashed line), and $\mathcal{F}_{gg}^{(6)}$ (dot-dashed line). The vertical dotted lines represent the saturation scale at the given value of $x_2$. In the figure, $Y=\ln \bigl(0.01/x_2\bigr)$. The plotted quantities do not include the factor $S_{\perp}/\alpha_s$ in Eq.~(\ref{eq:dipgluontmd}), common to all the gluon TMDs.}
\label{fig:2}
\end{figure}

We show in Fig.~\ref{fig:2} some of those gluon distributions for a target proton, as function of $k_t$, and for two values of $x_2$.
We do not show $\mathcal{F}_{gg}^{(2)}$ explicitly, but rather the difference $\mathcal{F}_{gg}^{(1)}-\mathcal{F}_{gg}^{(2)}$, which effectively plays a role in Eq.~(\ref{eq:full}).
The TMD distributions present three specific features, fully characterizing the dense component of our scattering.
Starting from the region where $k_t \gg 1$ GeV, we note that all the curves approach the same asymptotic behavior, i.e., an inverse power law, precisely equal to $k_t^{-2}$ at $x_2=x_0$ [panel (a)], with a smaller absolute slope after $x_2$ evolution [panel (b)].
As $k_t$ becomes of order 1 GeV, the TMD distributions start to separate, and quickly change their slope at a specific $k_t$, which corresponds approximately to the location of the maximum of  $F_{gg}^{(1)}-F_{gg}^{(2)}$.
This is the value of the saturation scale, $Q_s$, which we indicate in both panels with a vertical dotted line \footnote{More specifically, the maximum of $\mathcal{F}_{\rm adj}$ corresponds to the \textit{adjoint} saturation scale, which is 1.5 times bigger than $Q_s$, the \textit{fundamental} saturation scale which corresponds to the maximum of $F_{qg}^{(1)}$. This explains why the vertical line in Fig.~\ref{fig:2}(a) does not correspond to $Q_{s0}=\sqrt{0.2}$~GeV.}.
Note that the small-$x_2$ evolution has the effect of shifting the saturation scale to larger values.
Eventually, below the value of $Q_s$ the distributions become flat, and saturation is manifest. 
The difference $\mathcal{F}_{gg}^{(1)}-\mathcal{F}_{gg}^{(2)}$ goes to zero at $k_t=0$, consistently with Eq.~(\ref{Fadj}).

\section{The away-side peak from $\boldsymbol{\rm rc}$BK evolution: results and predictions}
\label{sec:4}
We can eventually employ the theoretical formalism introduced in the previous sections to compute azimuthal correlations of two hadrons in p+p and p+A collisions at $~\sqrt[]{s}=200~{\rm GeV}$.
We integrate Eq.~(\ref{eq:full}) over the momenta and rapidities of the produced hadrons, and study the behavior of the cross section as function of relative azimuthal angle, $\Delta\phi$.
In the notation of \cite{Albacete:2010pg}, the observable we want to calculate is
\begin{equation}
\label{eq:obs1}
N_{\rm pair}(\Delta\phi) = \int \frac{d\sigma^{pA\rightarrow \pi^0\pi^0X}}{d\Delta\phi~ dy_1~ dy_2~ dp^2_{t1}~ dp^2_{t2}} dy_1~ dy_2~ dp^2_{t1}~ dp^2_{t2}.
\end{equation}
The experimentally measured quantity is not directly given by Eq.~(\ref{eq:obs1}).
Experimentalists normalize $N_{\rm pair}(\Delta\phi)$ with the total number of hadrons that trigger the correlations, i.e.,
\begin{equation}
\label{eq:ntrigg}
N_{\rm trig} = \int \frac{d\sigma^{pA\rightarrow \pi^0 + X}}{dy~ dp^2_{t}}dy~dp^2_{t},
\end{equation}
in which we have introduced the cross section for single hadron production \cite{Albacete:2010bs}
\begin{align}
\label{eq:ntrig}
\nonumber \frac{d\sigma^{pA\rightarrow \pi^0 + X}}{dy~ dp^2_{t}} =  \int_{p_{t}\frac{e^{y}}{\sqrt[]{s}}}^1 \frac{dz}{z^2} \biggl\{ &\bigl[x_1 u(x_1,\mu^2)~D_{\pi^0/u}(z,\mu^2)+x_1 d(x_1,\mu^2)~D_{\pi^0/d}(z,\mu^2)\bigr] F(x_2,k_t) +\\
&+ x_1 g(x_1,\mu^2) D_{\pi^0/g}(x_2,\mu^2)~\tilde F(x_2,k_t) \biggr\},
\end{align}
where $F$ and $\tilde F$ are computed from the rcBK evolution equation as explained in the previous section.
The final observable is dubbed \textit{coincidence probability} by the STAR collaboration, and is given by
\begin{equation}
CP(\Delta\phi) = N_{\rm pair}(\Delta\phi) / N_{\rm trig}.
\end{equation}
Before showing our results, let us list all the details about the quantities needed in the calculation of $CP(\Delta\phi)$:
\begin{itemize}
\item The parton distribution functions (PDFs) describing the projectile are taken from the NLO MSTW2008 fits~\cite{Martin:2009iq};
\item The fragmentation functions (FFs) used are the recent DSS14 NLO sets~\cite{deFlorian:2014xna};
\item The strong coupling constant appearing in Eq.~(\ref{eq:full}) is calculated at NLO, and is given by the following expression
\begin{equation}
\alpha_s(\mu^2) = \frac{4\pi}{\bigr( 11-\frac{2}{3}N_f\bigl) \ln\bigl( \frac{\mu^2}{\Lambda^2}\bigr)},
\end{equation}
where we take $N_f=4$, and $\Lambda=197$~MeV. For $\mu^2$, we use the same scale employed in the PDFs and in the FFs (see item below);
\item The PDFs, the FFs, and $\alpha_s$ are computed at the scale $\mu^2=p_{t1}^2$, i.e., at the transverse momentum of the leading hadron.

\end{itemize}
\begin{figure}[t!]
\centering
\includegraphics[width=.7\linewidth]{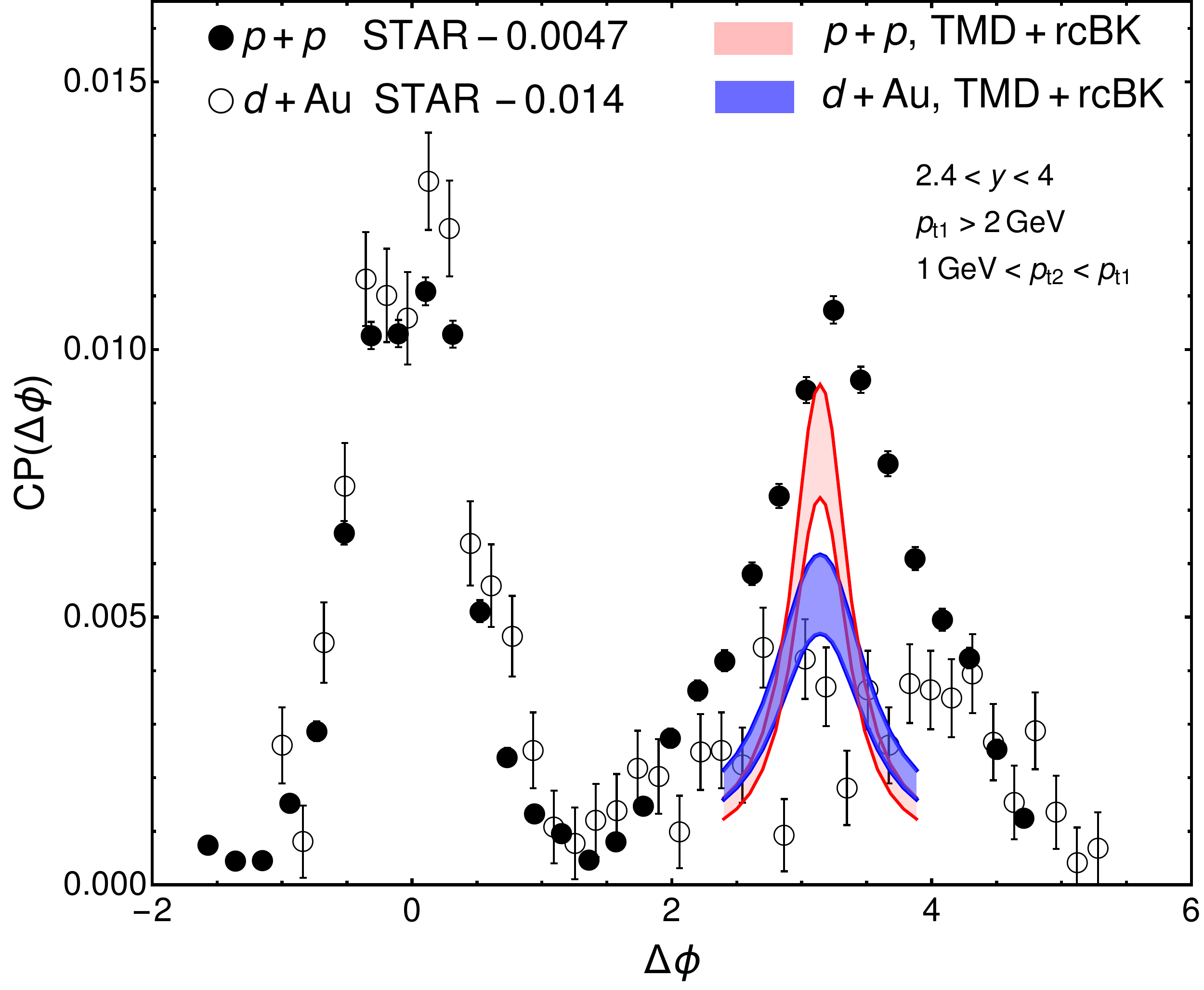}
\caption{The figure shows STAR data on azimuthal $\pi^0$ correlations at forward rapidity, in p+p collisions (circles) and central d+Au collisions (triangles) at $~\sqrt[]{s}=200~{\rm GeV}$. To remove fake two particle correlations which are essentially due to pileup effects, an arbitrary offset is added to push the STAR measurements close to 0 at the minimum of the correlation functions. Calculations of $CP(\Delta\phi)$ in our TMD+rcBK framework are shown as shaded bands. Light-shaded band: p+p collisions. Dark-shaded band: d+Au collisions. The meaning of the shaded bands is discussed in the text.}
\label{fig:3}
\end{figure}

\subsection{Comparison with Run--8 $\boldsymbol{d}$+Au RHIC data}
\begin{figure}[t!]
\centering
\includegraphics[width=.9\linewidth]{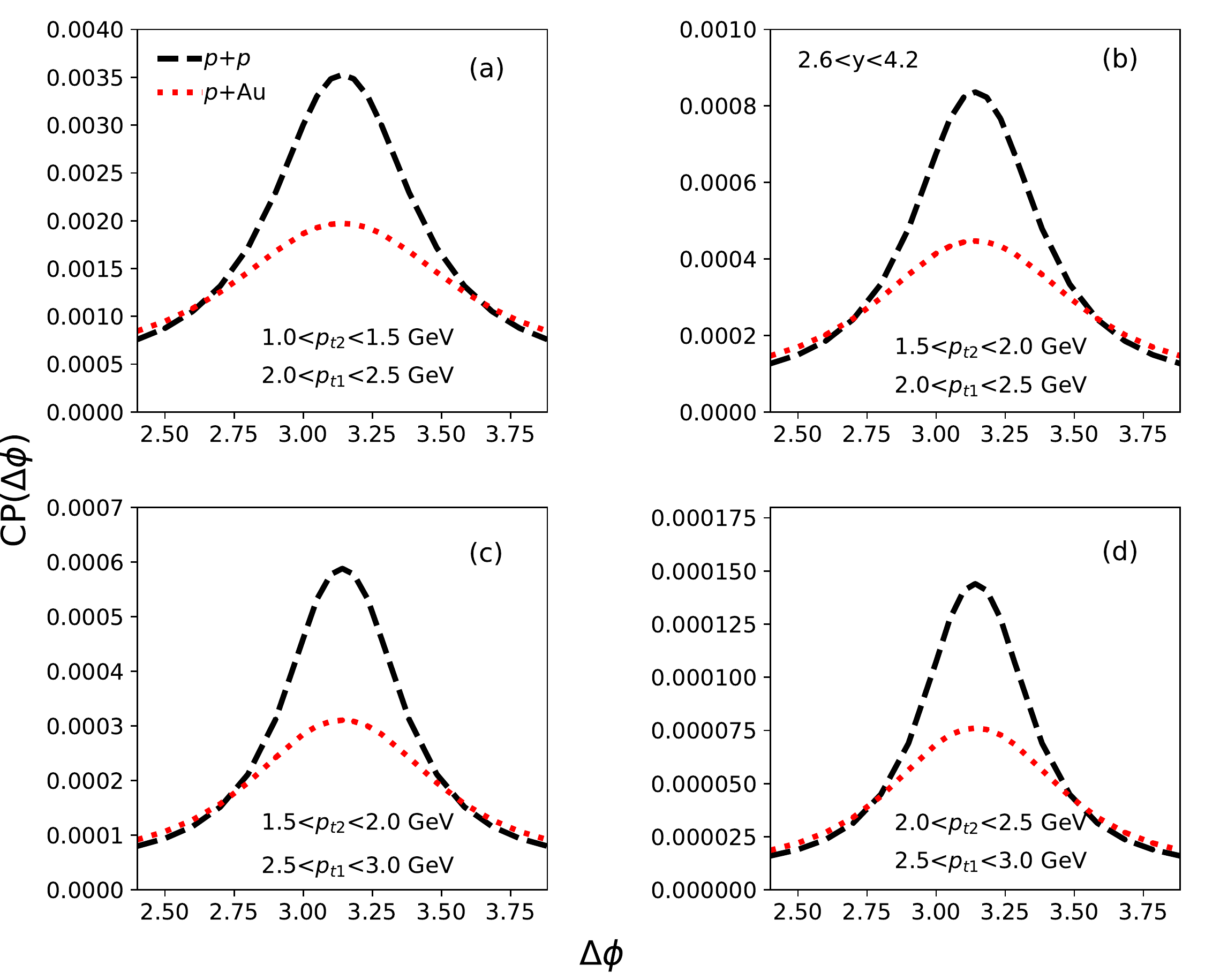}
\caption{In this figure we show predictions for azimuthal correlation of forward neutral pions in p+p (dashed line) and p+Au (dotted line) collisions at $~\sqrt[]{s}=200~{\rm GeV}$. Different panels correspond to different $p_t$ cuts applied to the cross section.}
\label{fig:4}
\end{figure}
Saturation effects are expected to yield a larger $CP(\Delta\phi)$ in $pp$ collisions than in $pA$ collisions, when $\Delta\phi$ is in the vicinity of $\pi$.
STAR data on $CP(\Delta\phi)$ \cite{Braidot:2010zh} for neutral pion correlations are shown as symbols in Fig.~\ref{fig:3}.
Data present a visible suppression of the correlation in d+Au collisions, suggesting that saturation effects may be effectively at play.
The outcome of integrating Eq.~(\ref{eq:full}) over the STAR kinematics in both p+p and $d$+Au collisions\footnote{Note that Eq.~(\ref{eq:full}) is suitable only for proton-nucleus collisions. A slightly different combinations of the PDFs and FFs at play is used to obtain the same cross section in the deuteron-nucleus case.} is shown as shaded bands, in the nearby of $\Delta\phi=\pi$.
The shaded bands represent the uncertainty in the choice of the factorization scale, $\mu^2$, in Eq.~(\ref{eq:full}).
The upper limit of the bands is obtained with $\mu^2=p_{t1}^2$, and the shaded area is obtained by taking a scale larger by 50\%.\footnote{We cannot test values of $\mu^2$ lower than $p_{t1}^2$, as they would lead to unreasonably small values of the factorization scale.}
In the following, we shall therefore always employ $\mu^2=p_{t1}^2$, which should provide the best agreement with data.

Figure~\ref{fig:3} shows that the suppression of the away-side peak provided by our calculation is in agreement with the data, although robust conclusions are impossible to draw due to the large uncertainties in d+Au collisions.
We also notice that our calculation reasonably captures the magnitude of $CP(\Delta\phi)$ at the away-side peak of p+p collisions.
What we fail in reproducing, though, is the width of the measured correlation in p+p, which appears to be broader than our result.
This has a simple explanation: in our calculation we are not supplementing the cross section with Sudakov factors, i.e., we do not take into account the radiation of soft gluons in both the initial and the final state, which would naturally provide a broadening of the correlation function. This was done in Ref.~\cite{bowen}, using GBW-type parametrization for the gluon TMDs, but it remains to be done with the rcBK gluon TMDs. An attempt of such Sudakov resummation in the case of di-jet production was made within the Kutak-Sapeta (KS) approach \cite{vanHameren:2014ala} (see Sec.~\ref{sec:5}), and the results are promising, in the sense that one observes a clear broadening around $\Delta\phi=\pi$ when Sudakov resummation is included. We finally note that the correlation function shown in Fig.~\ref{fig:3} is somewhat less flat than the one obtained in~\cite{Albacete:2010bs}. By comparison, our formalism is valid in a narrower window near $\Delta\phi=\pi$, but it is more accurate there.

\subsection{Predictions for $\boldsymbol{p}$+Au collisions}
In Fig.~\ref{fig:4} we present predictions for the away-side peak of neutral pions in p+p and p+Au collisions at $~\sqrt[]{s}=200$~GeV.
This is achieved by integrating Eq.~(\ref{eq:full}) over the kinematic cuts used by the STAR collaboration in their new analysis.
We predict that the away-side peak is suppressed in p+Au by a factor close to 2.
We find this conclusion to be rather independent of the $p_t$ window chosen for the measurement.

\section{Rapidity dependence of the suppression and comparison with the Kutak-Sapeta approach}
\label{sec:5}

A generic prediction of the CGC framework is that any effect due to gluon saturation should become less visible if we move towards more central rapidities, i.e., in our case, if we reduce the dilute-dense asymmetry by probing larger values of $x$ in the nuclei.
Consequently, the suppression of the away-side peak in p+A collisions relative to p+p should essentially fade away if we correlate particles in more central rapidity intervals.
It is important to stress that the dependence on rapidity is a very specific feature of the saturation framework, which is not predicted by typical competing effects, e.g., conservation of total transverse momentum~\cite{Borghini:2003ur}, or other energy-momentum conservation corrections which are relevant in the proximity of $x_1\rightarrow1$~\cite{Kopeliovich:2005ym,Frankfurt:2007rn}.
Another competing description is that reported by Kang \textit{et al.}~\cite{Kang:2011bp}, who manage to describe the suppression of the away-side peak without resorting to a CGC description, but solely from (cold) nuclear transverse-momentum broadening effects.
Such models do not predict a specific dependence on the rapidity, so that the CGC interpretation would be strongly favored if such dependence is observed in data.
Let us stress that the away-side peak in different rapidity intervals could be easily measured at the STAR or at the LHCb detectors, which present wide rapidity coverages.

Let us show, then, what our formalism predicts for the rapidity dependence of the suppression of the away-side peak.
For reasons which will appear clear in the following discussion, it is very instructive to perform calculations and show results using both our rcBK formalism and the alternative Kutak-Sapeta (KS) approach \cite{Kutak:2012rf}, which we briefly review below.

\begin{figure}[t!]
\centering
\includegraphics[width=\linewidth]{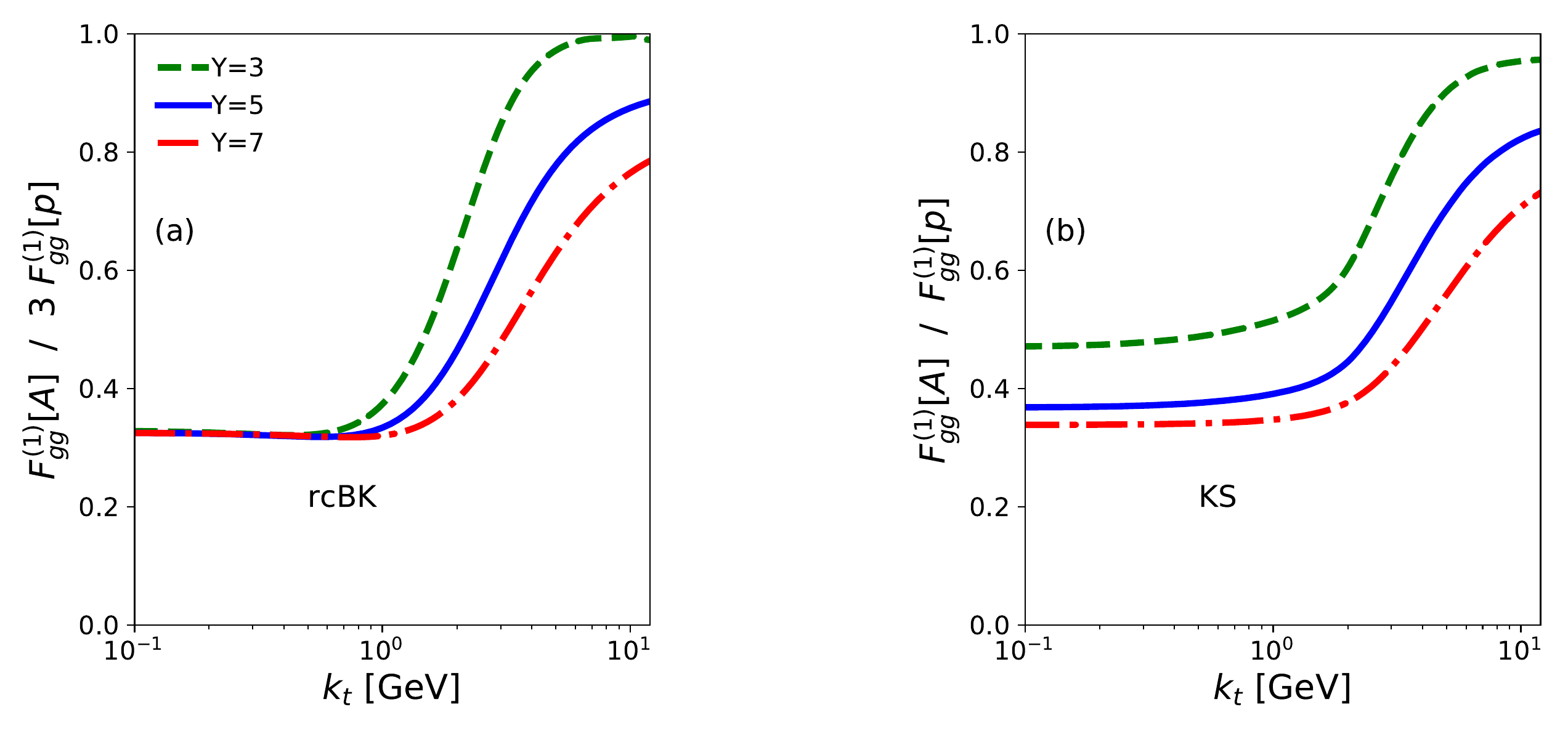}
\caption{The figure shows $\mathcal{F}_{gg}^{(1)}$ for a target nucleus divided by the same quantity for a target proton, as function of $k_t$. Results are shown within two different evolution schemes, namely rcBK [panel (a)] and KS approximation [panel(b)]. The ratio is taken at different values of $x_2$, indicated with different line styles. In the figure, $Y=\ln \bigl(0.01/x_2\bigr)$.}
\label{fig:5}
\end{figure}

In the KS approach, the momentum space version of the BK equation is used (written below for ${\cal F}_{p}=\pi\mathcal{F}_{qg}^{(1)}$, for a target proton):
\begin{multline} 
{\cal F}_p(x,k^2) \; = \; {\cal F}^{(0)}_p(x,k^2) \\
+\,\frac{\alpha_s N_c}{\pi}\int_x^1 \frac{dz}{z} \int_{\mu^2}^{\infty}
\frac{dl^2}{l^2} \,   \bigg\{ \, \frac{l^2{\cal F}_p(\frac{x}{z},l^2)\,   -\,
k^2{\cal F}_p(\frac{x}{z},k^2)}{|l^2-k^2|}   +\,
\frac{k^2{\cal F}_p(\frac{x}{z},k^2)}{|4l^4+k^4|^{\frac{1}{2}}} \,
\bigg\} \\
-\frac{2\alpha_s^2}{R^2}\left[\left(\int_{k^2}^{\infty}\frac{dl^2}{l^2}{\cal F}_p(x,l^2)\right)^2
+
{\cal F}_p(x,k^2)\int_{k^2}^{\infty}\frac{dl^2}{l^2}\ln\left(\frac{l^2}{k^2}\right){\cal F}_p(x,l^2)\right]\ .
\label{eq:fkovresKS} 
\end{multline}
This way of writing the BK equation is convenient as it allows to include relatively easily some higher-order corrections, and in particular running-coupling corrections \cite{Kutak:2003bd}. To write down the non-linear term of Eq.~(\ref{eq:fkovresKS}) (last line in the equation) for the impact-parameter-integrated gluon distribution, it is assumed that integration over impact parameter yields $\int d^2b=\pi R^2$, where $R$ is the radius of the target proton.
The evolution of the gluon TMD in the case of a nucleus, ${\cal F}_A$, is then obtained through the following formal substitution in Eq.~(\ref{eq:fkovresKS}),
\begin{equation}
\frac{1}{R^2} \to  c\, \frac{A}{R_A^2}\,,
\qquad {\rm where} \quad
R^2_{\text{A}}= R^2\, A^{2/3} \,.
\label{eq:radius}
\end{equation}
In the above equation, $R_A$ is the nuclear radius, $A$ is the mass number ($A=208$ for Pb), and $c$ is a parameter that is supposed to vary between 0.5 and 1, to assess the uncertainty related to the nonlinear term. 
The density ${\cal F}_A$ obtained from Eq.~(\ref{eq:fkovresKS}) with the
substitution above is the nuclear gluon density normalized to the number of nucleons in the nuclei. 

The KS evolution in Eq.~(\ref{eq:fkovresKS}) is $A$-dependent through the non-linear term (it has to be so, since ${\cal F}_A$ is an impact parameter integrated distribution), but the prescription for the initial condition is to choose the same in the nuclear case as in the proton case, i.e., ${\cal F}_A(x_0,k^2)={\cal F}_p(x_0,k^2)$. This is the major difference with respect to the approach presented in Sec.~\ref{sec:3}, where an $A$-dependent initial condition and an $A$-independent evolution were used.

Figure~\ref{fig:5} shows an illustration of the effect due to this difference between the rcBK and the KS approaches, which are both based on the same small-x evolution. In the figure we show $\mathcal{F}_{gg}^{(1)}$ for a target nucleus divided by the same quantity for a target proton\footnote{The factor 3 appearing in the denominator of the rcBK ratio corresponds to the initial value of the ratio $Q_s^2 {\rm[A]}/Q_s^2 {\rm[pp]}$, as introduced in Sec.~\ref{sec:3}.}, for different values of rapidity $Y$, which is defined as $Y=\ln \bigl( 0.01/x_2 \bigr)$.
On the left, the rcBK distributions predict the same amount of suppression at each value of $Y$ in the fully saturated region, $k_t\sim0$, because the small-$x$ evolution is $A$-independent.
This is not the case in panel (b), where the ratio in the KS scheme is equal to unity for $Y=0$ (not shown), and the difference in the evolution of the nucleus with respect to the proton is manifest already at $k_t=0$.
Note that the plot is drawn for $c=0.5$, but we stress that the qualitative picture is essentially independent of the choice of this constant.
\begin{figure}[t!]
\centering
\includegraphics[width=\linewidth]{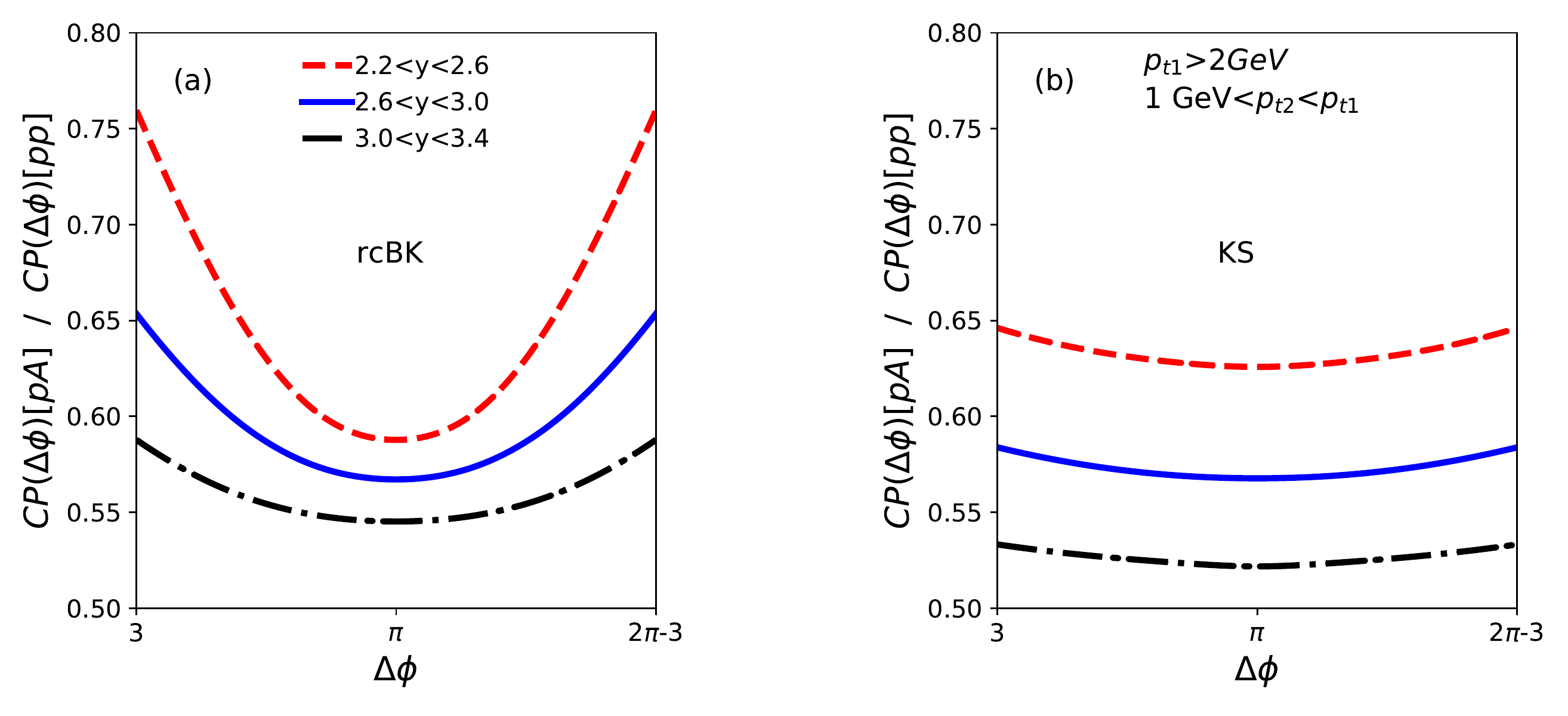}
\caption{The figure shows the ratio $CP(\Delta\phi)_{pA}/CP(\Delta\phi)_{pp}$ around $\Delta\phi=\pi$. Different line styles represent different rapidity intervals. Panel (a) shows results with gluon TMDs obtained as described in section \ref{sec:3}. In panel (b) the TMDs are obtained using the KS scheme, with $c=0.5$.}
\label{fig:6}
\end{figure}

This difference shown in Fig.~\ref{fig:5} has a non-negligible impact on the rapidity dependence of the suppression of the away-side peak, which is the subject under study in this section.
To show this, we calculate the ratio $CP(\Delta\phi)$ in p+Au over the same quantity in p+p using both the standard rcBK approach and the KS alternative proposal\footnote{The KS implementaion we have in mind in not directly the one performed in \cite{Kutak:2012rf} which involves a single gluon TMD, but rather an adaptation of it to the away-side peak region, involving the several gluon TMDs needed just as in \eqref{eq:full}.}, and we look at its dependence with rapidity.
We keep the old STAR kinematics of Fig.~\ref{fig:3} for the $p_t$ of the produced hadrons, and we compute $CP(\Delta\phi)_{pA}/CP(\Delta\phi)_{pp}$, with an obvious meaning of the notation, around $\Delta\phi=\pi$ in different intervals of rapidity.
Results are shown in Fig.~\ref{fig:6}.
We find that both schemes provide a hierarchy as function of rapidity expected in the saturation framework: To more central rapidities correspond larger values of the ratio around $\Delta\phi=\pi$, i.e., less suppression at the away-side peak.
We stress that this is a peculiar feature of the saturation framework, and we strongly encourage measurements of this ratio in different rapidity intervals, which could provide, arguably, the strongest possible evidence in favor of the saturation picture.
In addition, we expect such quantity to be almost unaffected by the uncertainties on the factorization scale (which turned out to be quite large in Fig.~\ref{fig:3}), as they are likely to cancel in the ratio.

Besides confirming the generic prediction of the CGC framework, precise measurements in p+Au collisions might as well shed light on the very validity of the approaches taken for the small-$x$ evolution of the dense targets.
In Fig.~\ref{fig:6} we observe two notable differences between rcBK and KS.
First, the dependence on rapidity at $\Delta\phi=\pi$ is about twice stronger in the KS approach [panel (b)]: This results from having a small-$x$ evolution at low $k_t$ (Fig.~\ref{fig:5}).
Second, the rcBK case presents ratios which grow towards unity much faster as we move away from the back-to-back region.
Specifically, the ratio at $\Delta\phi=3$ is larger by 15\% in the rcBK scheme.
Such visible differences are expected to be sizable in the upcoming data, and would help improve significantly our understanding of the evolution equations of QCD in the nonlinear small-$x$ regime.

\section{conclusion}
\label{sec:6}
We have calculated the production of back-to-back pions in p+A and p+p collisions at RHIC energies, using the state-of-the-art CGC framework, i.e., the cross section reported in Eq.~\eqref{eq:full}. We have developed a novel approach for the small-$x$ evolution of the TMD gluon distributions ${\cal F}_{ag}^{(i)}$, in which they are obtained from the BK evolution with an evolution kernel that includes running coupling corrections.
The evolution is identical for proton and nuclear targets, the only difference being the value of $Q_{s}^2$ at the initial condition.
The validity of our framework is confirmed by the good agreement observed between the available data and our results in Fig.~\ref{fig:3}.

We thus derived genuine predictions of the CGC theory. 
The away-side peak in upcoming p+Au data is suppressed by about a factor 2 with respect to p+p collisions (Fig.~\ref{fig:4}), and this suppression tends to disappear as we reduce the dilute-dense asymmetry of the problem (Fig.~\ref{fig:6}).
We stress, once more, that the combination of these two effects is a much stronger probe of gluon saturation than the suppression of the away-side peak alone. 
We have further compared the expectation of our framework to those of another state-of-the-art rcBK implementation, namely, the KS approach.
Using the observable proposed in Fig.~\ref{fig:6}, p+Au data will potentially allow us to make a data-driven distinction between these two schemes of small-$x$ evolution.

Before concluding, we stress that our calculation lacks an important ingredient: The inclusion of the soft gluon resummation, i.e., of Sudakov factors attached to the cross section which could potentially solve our problem of a too narrow correlation peak around $\Delta\phi=\pi$ (Fig.~\ref{fig:3}). This improvement of our formalism will be presented in an upcoming publication.

\section{Acknowledgements}
The work of CM was supported in part by the Agence Nationale de la Recherche under the project ANR-16-CE31-0019-02. The work of MM has been supported by the grant 17-04505S of the Czech Science Foundation (GACR) and has been performed in the framework of COST Action CA15213 THOR. Computational resources were provided by the CESNET LM2015042 and the CERIT Scientific Cloud LM2015085, provided under the program ``Projects of Large Research, Development, and Innovations Infrastructures''.


\begin{thebibliography}{99}
\bibitem{Gelis:2010nm}
  F.~Gelis, E.~Iancu, J.~Jalilian-Marian and R.~Venugopalan,
  Ann.\ Rev.\ Nucl.\ Part.\ Sci.\  {\bf 60} (2010) 463
  doi:10.1146/annurev.nucl.010909.083629
  [arXiv:1002.0333 [hep-ph]].

\bibitem{Gelis:2012ri} 
  F.~Gelis,
  Int.\ J.\ Mod.\ Phys.\ A {\bf 28}, 1330001 (2013)
  doi:10.1142/S0217751X13300019
  [arXiv:1211.3327 [hep-ph]].


\bibitem{Blaizot:2016qgz} 
  J.~P.~Blaizot,
  Rept.\ Prog.\ Phys.\  {\bf 80}, no. 3, 032301 (2017)
  doi:10.1088/1361-6633/aa5435
  [arXiv:1607.04448 [hep-ph]].


\bibitem{Marquet:2007vb} 
  C.~Marquet,
  Nucl.\ Phys.\ A {\bf 796}, 41 (2007)
  doi:10.1016/j.nuclphysa.2007.09.001
  [arXiv:0708.0231 [hep-ph]].


\bibitem{Tuchin:2009nf} 
  K.~Tuchin,
  Nucl.\ Phys.\ A {\bf 846}, 83 (2010)
  doi:10.1016/j.nuclphysa.2010.06.001
  [arXiv:0912.5479 [hep-ph]].


\bibitem{Albacete:2010pg} 
  J.~L.~Albacete and C.~Marquet,
  Phys.\ Rev.\ Lett.\  {\bf 105}, 162301 (2010)
  doi:10.1103/PhysRevLett.105.162301
  [arXiv:1005.4065 [hep-ph]].


\bibitem{Stasto:2011ru} 
  A.~Stasto, B.~W.~Xiao and F.~Yuan,
  Phys.\ Lett.\ B {\bf 716}, 430 (2012)
  doi:10.1016/j.physletb.2012.08.044
  [arXiv:1109.1817 [hep-ph]].


\bibitem{Lappi:2012nh} 
  T.~Lappi and H.~Mantysaari,
  Nucl.\ Phys.\ A {\bf 908}, 51 (2013)
  doi:10.1016/j.nuclphysa.2013.03.017
  [arXiv:1209.2853 [hep-ph]].


\bibitem{Ayala:2016lhd} 
  A.~Ayala, M.~Hentschinski, J.~Jalilian-Marian and M.~E.~Tejeda-Yeomans,
  Phys.\ Lett.\ B {\bf 761}, 229 (2016)
  doi:10.1016/j.physletb.2016.08.035
  [arXiv:1604.08526 [hep-ph]].


\bibitem{vanHameren:2016ftb} 
  A.~van Hameren, P.~Kotko, K.~Kutak, C.~Marquet, E.~Petreska and S.~Sapeta,
  JHEP {\bf 1612}, 034 (2016)
  doi:10.1007/JHEP12(2016)034
  [arXiv:1607.03121 [hep-ph]].


\bibitem{Kotko:2017oxg} 
  P.~Kotko, K.~Kutak, S.~Sapeta, A.~M.~Stasto and M.~Strikman,
  Eur.\ Phys.\ J.\ C {\bf 77}, no. 5, 353 (2017)
  doi:10.1140/epjc/s10052-017-4906-6
  [arXiv:1702.03063 [hep-ph]].


\bibitem{Albacete:2014fwa} 
  J.~L.~Albacete and C.~Marquet,
  Prog.\ Part.\ Nucl.\ Phys.\  {\bf 76}, 1 (2014)
  doi:10.1016/j.ppnp.2014.01.004
  [arXiv:1401.4866 [hep-ph]].


\bibitem{Braidot:2010zh} 
  E.~Braidot [STAR Collaboration],
  arXiv:1005.2378 [hep-ph].


\bibitem{Adare:2011sc} 
  A.~Adare {\it et al.} [PHENIX Collaboration],
  Phys.\ Rev.\ Lett.\  {\bf 107}, 172301 (2011)
  doi:10.1103/PhysRevLett.107.172301
  [arXiv:1105.5112 [nucl-ex]].


\bibitem{JalilianMarian:2004da} 
  J.~Jalilian-Marian and Y.~V.~Kovchegov,
  Phys.\ Rev.\ D {\bf 70}, 114017 (2004)
  Erratum: [Phys.\ Rev.\ D {\bf 71}, 079901 (2005)]
  doi:10.1103/PhysRevD.71.079901, 10.1103/PhysRevD.70.114017
  [hep-ph/0405266].

\bibitem{Kutak:2012rf}
  K.~Kutak and S.~Sapeta,
  Phys.\ Rev.\ D {\bf 86} (2012) 094043
  doi:10.1103/PhysRevD.86.094043
  [arXiv:1205.5035 [hep-ph]].

\bibitem{Dominguez:2010xd}
  F.~Dominguez, B.~W.~Xiao and F.~Yuan,
  Phys.\ Rev.\ Lett.\  {\bf 106} (2011) 022301
  doi:10.1103/PhysRevLett.106.022301
  [arXiv:1009.2141 [hep-ph]].

\bibitem{Dominguez:2011wm} 
  F.~Dominguez, C.~Marquet, B.~W.~Xiao and F.~Yuan,
  Phys.\ Rev.\ D {\bf 83}, 105005 (2011)
  doi:10.1103/PhysRevD.83.105005
  [arXiv:1101.0715 [hep-ph]].


\bibitem{Marquet:2016cgx} 
  C.~Marquet, E.~Petreska and C.~Roiesnel,
  JHEP {\bf 1610}, 065 (2016)
  doi:10.1007/JHEP10(2016)065
  [arXiv:1608.02577 [hep-ph]].


\bibitem{Marquet:2017xwy} 
  C.~Marquet, C.~Roiesnel and P.~Taels,
  Phys.\ Rev.\ D {\bf 97}, no. 1, 014004 (2018)
  doi:10.1103/PhysRevD.97.014004
  [arXiv:1710.05698 [hep-ph]].


\bibitem{Boer:2017xpy} 
  D.~Boer, P.~J.~Mulders, J.~Zhou and Y.~j.~Zhou,
  JHEP {\bf 1710}, 196 (2017)
  doi:10.1007/JHEP10(2017)196
  [arXiv:1702.08195 [hep-ph]].


\bibitem{Altinoluk:2018uax} 
  T.~Altinoluk, N.~Armesto, A.~Kovner, M.~Lublinsky and E.~Petreska,
  arXiv:1802.01398 [hep-ph].


\bibitem{Petreska:2018cbf} 
  E.~Petreska,
  arXiv:1804.04981 [hep-ph].


\bibitem{Kotko:2015ura} 
  P.~Kotko, K.~Kutak, C.~Marquet, E.~Petreska, S.~Sapeta and A.~van Hameren,
  JHEP {\bf 1509}, 106 (2015)
  doi:10.1007/JHEP09(2015)106
  [arXiv:1503.03421 [hep-ph]].

\bibitem{Albacete:2010sy}
  J.~L.~Albacete, N.~Armesto, J.~G.~Milhano, P.~Quiroga-Arias and C.~A.~Salgado,
  Eur.\ Phys.\ J.\ C {\bf 71} (2011) 1705
  doi:10.1140/epjc/s10052-011-1705-3
  [arXiv:1012.4408 [hep-ph]].

\bibitem{Balitsky:1995ub} 
  I.~Balitsky,
  Nucl.\ Phys.\ B {\bf 463}, 99 (1996)
  doi:10.1016/0550-3213(95)00638-9
  [hep-ph/9509348].


\bibitem{Kovchegov:1999yj} 
  Y.~V.~Kovchegov,
  Phys.\ Rev.\ D {\bf 60}, 034008 (1999)
  doi:10.1103/PhysRevD.60.034008
  [hep-ph/9901281].


\bibitem{Albacete:2007yr} 
  J.~L.~Albacete and Y.~V.~Kovchegov,
  Phys.\ Rev.\ D {\bf 75}, 125021 (2007)
  doi:10.1103/PhysRevD.75.125021
  [arXiv:0704.0612 [hep-ph]].


\bibitem{Albacete:2010bs} 
  J.~L.~Albacete and C.~Marquet,
  Phys.\ Lett.\ B {\bf 687}, 174 (2010)
  doi:10.1016/j.physletb.2010.02.073
  [arXiv:1001.1378 [hep-ph]].


\bibitem{Fujii:2006ab} 
  H.~Fujii, F.~Gelis and R.~Venugopalan,
  Nucl.\ Phys.\ A {\bf 780}, 146 (2006)
  doi:10.1016/j.nuclphysa.2006.09.012
  [hep-ph/0603099].


\bibitem{Kovchegov:2008mk} 
  Y.~V.~Kovchegov, J.~Kuokkanen, K.~Rummukainen and H.~Weigert,
  Nucl.\ Phys.\ A {\bf 823}, 47 (2009)
  doi:10.1016/j.nuclphysa.2009.03.006
  [arXiv:0812.3238 [hep-ph]].


\bibitem{Marquet:2010cf} 
  C.~Marquet and H.~Weigert,
  Nucl.\ Phys.\ A {\bf 843}, 68 (2010)
  doi:10.1016/j.nuclphysa.2010.05.056
  [arXiv:1003.0813 [hep-ph]].


\bibitem{Dumitru:2011vk} 
  A.~Dumitru, J.~Jalilian-Marian, T.~Lappi, B.~Schenke and R.~Venugopalan,
  Phys.\ Lett.\ B {\bf 706}, 219 (2011)
  doi:10.1016/j.physletb.2011.11.002
  [arXiv:1108.4764 [hep-ph]].


\bibitem{Iancu:2011nj} 
  E.~Iancu and D.~N.~Triantafyllopoulos,
  JHEP {\bf 1204}, 025 (2012)
  doi:10.1007/JHEP04(2012)025
  [arXiv:1112.1104 [hep-ph]].


\bibitem{Alvioli:2012ba} 
  M.~Alvioli, G.~Soyez and D.~N.~Triantafyllopoulos,
  Phys.\ Rev.\ D {\bf 87}, no. 1, 014016 (2013)
  doi:10.1103/PhysRevD.87.014016
  [arXiv:1212.1656 [hep-ph]].


\bibitem{Martin:2009iq} 
  A.~D.~Martin, W.~J.~Stirling, R.~S.~Thorne and G.~Watt,
  Eur.\ Phys.\ J.\ C {\bf 63}, 189 (2009)
  doi:10.1140/epjc/s10052-009-1072-5
  [arXiv:0901.0002 [hep-ph]].


\bibitem{deFlorian:2014xna} 
  D.~de Florian, R.~Sassot, M.~Epele, R.~J.~Hernández-Pinto and M.~Stratmann,
  Phys.\ Rev.\ D {\bf 91}, no. 1, 014035 (2015)
  doi:10.1103/PhysRevD.91.014035
  [arXiv:1410.6027 [hep-ph]].


\bibitem{bowen}
A.~Stasto, S-Y.~Wei, B-W~ Xiao and F.~Yuan, to appear

\bibitem{vanHameren:2014ala}
  A.~van Hameren, P.~Kotko, K.~Kutak and S.~Sapeta,
  Phys.\ Lett.\ B {\bf 737} (2014) 335
  doi:10.1016/j.physletb.2014.09.005
  [arXiv:1404.6204 [hep-ph]].

\bibitem{Borghini:2003ur} 
  N.~Borghini,
  Eur.\ Phys.\ J.\ C {\bf 30}, 381 (2003)
  doi:10.1140/epjc/s2003-01265-6
  [hep-ph/0302139].


\bibitem{Kopeliovich:2005ym} 
  B.~Z.~Kopeliovich, J.~Nemchik, I.~K.~Potashnikova, M.~B.~Johnson and I.~Schmidt,
  Phys.\ Rev.\ C {\bf 72}, 054606 (2005)
  doi:10.1103/PhysRevC.72.054606
  [hep-ph/0501260].


\bibitem{Frankfurt:2007rn} 
  L.~Frankfurt and M.~Strikman,
  Phys.\ Lett.\ B {\bf 645}, 412 (2007)
  doi:10.1016/j.physletb.2007.01.007
  [nucl-th/0603049].


\bibitem{Kang:2011bp} 
  Z.~B.~Kang, I.~Vitev and H.~Xing,
  Phys.\ Rev.\ D {\bf 85}, 054024 (2012)
  doi:10.1103/PhysRevD.85.054024
  [arXiv:1112.6021 [hep-ph]].


\bibitem{Kutak:2003bd} 
  K.~Kutak and J.~Kwiecinski,
  Eur.\ Phys.\ J.\ C {\bf 29}, 521 (2003)
  doi:10.1140/epjc/s2003-01236-y
  [hep-ph/0303209].
 
\end{thebibliography}
\end{document}